Title:

# Non-Transitivity of the Win Ratio and the Area Under the Receiver Operating Characteristics Curve (AUC): a case for evaluating the strength of stochastic comparisons.

Demler, Olga V.[1,2], Demler, Ilona A.[3]

[1]Brigham & Women's Hospital, Boston, MA, USA
[2]ETH Zurich, Zurich, Switzerland
[3]California Institute of Technology, Pasadena, CA, USA

## Abstract

The win ratio (WR) is a novel statistic used in randomized controlled trials that can account for hierarchies within event outcomes. In this paper we report and study the long-run non-transitive behavior of the win ratio and the closely related Area Under the Receiver Operating Characteristics Curve (AUC) and argue that their transitivity cannot be taken for granted. Crucially, traditional within-group statistics (i.e., comparison of means) are always transitive, while the WR can detect non-transitivity. Non-transitivity provides valuable information on the stochastic relationship between two treatment groups, which should be tested and reported. We specify the necessary conditions for transitivity, the sufficient conditions for non-transitivity, and demonstrate non-transitivity in a real-life large randomized controlled trial for the WR of time-to-death. Our results can be used to rule out or evaluate the possibility of non-transitivity and show the importance of studying the strength of stochastic relationships.

1. Introduction

The win ratio is a novel and ubiquitous measure of treatment efficacy in randomized controlled trials. The win ratio (Finkelstein and Schoenfeld 1999, Pocock et al. 2012) summarizes pairwise comparisons of participants from two treatment arms of a trial in terms of their times to events. A treatment "wins" over the placebo if for a given pair of participants, the time to event for participant $i$ from the treatment arm (arm A) is longer than the time to event for a participant $j$ from the placebo arm (arm B) ($T_{Ai} > T_{Bj}$). The win ratio is the ratio of the number of wins for a treatment arm to the number of wins for a placebo arm. A win ratio greater than 1 is interpreted as the treatment winning over the placebo. In many clinical trials, the primary endpoint is composed of several types of events. The final analysis considers only the time to the first even, however, fatal and severe non-fatal events need to be prioritized instead. The win ratio can accommodate more than one type of outcome as well as their hierarchy, e.g., a primary outcome (death) can take precedence over a secondary outcome (non-fatal event). Thus, the win ratio uses more nuanced outcome information which translates into more power (Ferreira et al. 2020). The win ratio when no hierarchy is integrated has a direct relationship with the hazard ratio (De Neve and Gerds 2020). These attractive properties explain why the win ratio is already used in the main publications of several large randomized controlled trials in cardiology, neurology, and pulmonology (Solomon et al. 2022, Novack et al. 2020, Benatar et al. 2018, Maurer et al. 2018). As a novel statistic, the win ratio is still in development and poses several interesting questions. Exploring the properties of the win ratio continues to be an active, ongoing area of research.

**1.1 New contribution**

In this paper we report that non-transitivity is one of the properties of the long-run behavior of the win ratio and study it in detail. Critically, in the medical context a non-transitive loop implies an ambiguity in treatment efficacy. There are several studies of transitivity and its opposite, non-transitivity, in game theory, theoretical statistics, econometrics, and quantum physics. In these fields, non-transitivity is defined as a property of distributions of random variables, i.e. a property of their long-run behavior. A famous example of non-transitivity is the set of four non-transitive Efron dice A, B C, D(Gardner 1970). The scores on the dice are distributed in such a way that on average, die A beats die B, B beats C, C beats D, and D paradoxically beats A. In this example, we say that die A wins over die B if the numbers it rolls are higher in the long run, i.e. when Pr(a>b)>1/2. This Pr(a>b) metric is also one of the interpretations of the Area Under the Receiver Characteristics Curve (AUC) (Bamber and McNeil 1975), a ubiquitous statistic that evaluates the discrimination ability of a classifier in machine learning or risk prediction model. Thus, the long-term outcome of a game of dice relies on measuring their pairwise AUC. The Efron dice example demonstrates that the AUC can be non-transitive. We can write Pr(a>b)>1/2 as $A \prec_{AUC} B$ and the non-transitive loop of Efron dice as $A \prec_{AUC} B \prec_{AUC} C \prec_{AUC} D \prec_{AUC} A$.

The AUC is also linked to the win ratio via a monotonic transformation. Therefore the win ratio can be non-transitive in the long run. In our work, we find non-transitive win ratios in a large randomized controlled trial when comparing time to death (Green et al. 2015, Bertagnolli et al. 2017). Specifically, we calculate the win ratios for four randomized treatment groups (A, B, C and D). We find that the win ratios of time to all-cause death produce non-transitive win ratios. This means that on average, participants from arm A live longer than those in arm B, those from B longer than from arm C, from C longer than from arm D, and paradoxically from D longer than from arm A. Here, "longer" is defined as having a win ratio of greater than 1 (Figure 1). Or $A \prec_{WR} B \prec_{WR} C \prec_{WR} D \prec_{WR} A$. This real-life example of a large randomized controlled trial confirms that win ratios can form non-transitive loops. In this paper we will show that this is a property of the long-run behavior of certain distributions of underlying random variables (in this example the time to death), meaning that larger sample sizes will not ensure transitivity. While the example in Figure 1 has very weak win ratios, we will also show that win ratios of any strength can be part of a non-transitive loop.

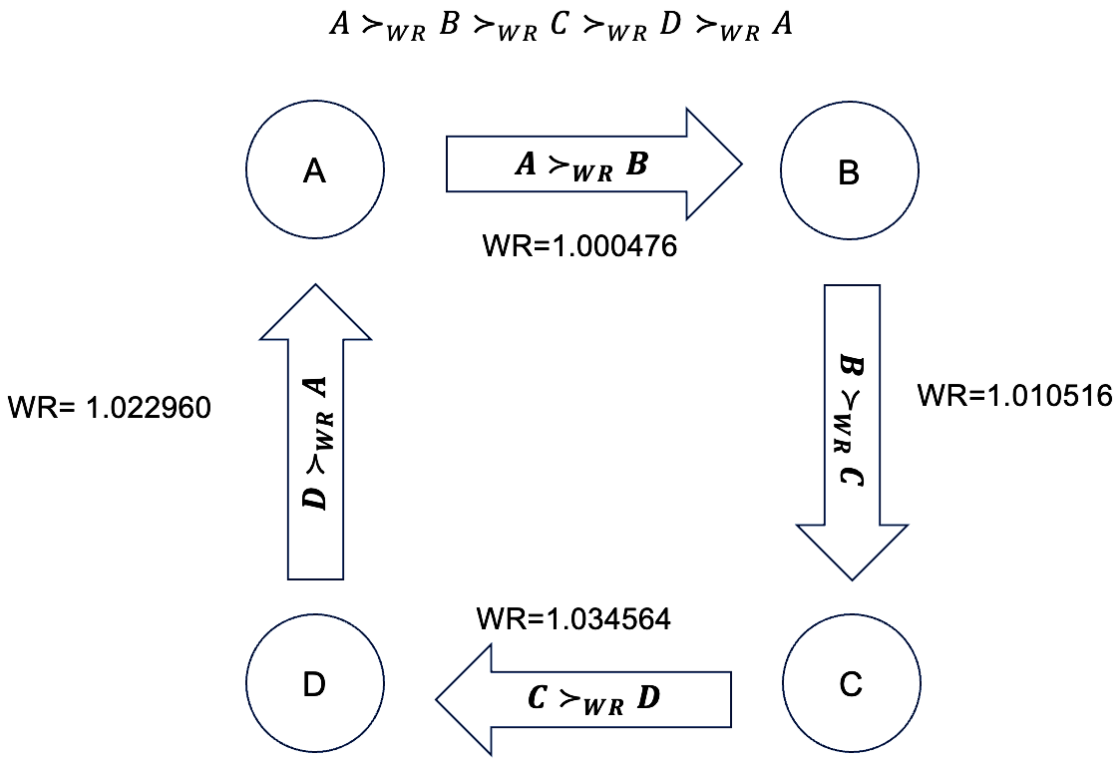

Figure 1. Non-transitive loop observed in a large randomized controlled trial (N=25,371). The win ratios were calculated by comparing the time to all-cause death for pairs of randomized treatment arms (A, B, C and D). The person-level data for this trial is available for download from project Data Sphere (Bertagnolli et al. 2017, Green et al. 2015). More information on how the win ratio was calculated in the presence of censoring is presented in the Appendix.

We should prefer transitive treatments $\alpha, \beta, \gamma$ over non-transitive treatments A, B, C, if the traditional measures of efficacy for both treatments $\alpha, \beta, \gamma$ and A, B, C are similar. We argue that metrics that can detect non-transitivity should be reported, while non-transitivity as a property should be acknowledged and studied. To our knowledge, non-transitivity of the win ratio in the classical "Efron" sense, as a property of its long-run behavior, has not been reported and studied. It is therefore important to study the conditions under which treatments can form non-transitive loops, the conditions that rule out non-transitivity, and the implications of these findings for the medical field.

## 1.2 Organization of the paper

This paper is organized as follows. We prove the existence of non-transitive loops by providing examples of distributions for which the AUC and win ratios exhibit non-transitive behavior in the long run. We provide sufficient conditions on the distribution functions that guarantee transitivity of the WR and AUC for any number of comparisons. Then we focus on non-transitive loops of length three. We use the results of Trybula (Trybuła 1961, 1965) to fully define a subspace of AUCs and WRs of all possible non-transitive loops of length three. This allows us to provide thresholds on a pair of AUCs and WRs that guarantee that together they can only comprise transitive comparisons. We also report the necessary conditions for non-transitivity. Finally, we discuss the implications of these findings for medical research in the broader context of the strength of stochastic comparisons.

## 2. Definitions

### 2.1 Win Ratio (WR)

Let's denote as $x_i, y_j, z_k$ as the time from baseline to event for participants $i$ and $j$ in a randomized controlled trial in treatment arms 1, 2, and 3. Suppose that the trial consists of three treatment arms. To calculate the win ratio of arm 1 compared to arm 2, we form all possible pairs $(i,j)$ of participants, where $i,j$ are participants from arms 1 and 2 correspondingly. Assuming no ties in times to event, the WR is defined as: $WR_{12} = \frac{Pr[x_i > y_j]}{Pr[x_i < y_j]}$ and can be estimated as $\widehat{WR_{12}} = \frac{\sum I[x_i > y_j]}{\sum I[x_i < y_j]}$, where $I[.]$ is an indicator function. $I[T_{1i} > T_{2j}] = 1$ when treatment 1 "wins" over treatment 2. Therefore, $WR_{12}$ is the ratio of the number of wins of treatment 1 to the number of wins of treatment 2. $WR_{12} > 1.0$ means that treatment 1 is beneficial when compared to treatment 2. To test $H_0: WR_{12} > 1.0$ we can use the equivalence of WR and Mann-Whitney statistics to use the asymptotic properties of the latter (Dong et al. 2019). In this study we assume that participants $i,j$ are independent for any $i \neq j$.

### 2.2 AUC

The AUC is defined as $Pr[x_i < y_j]$. Therefore $\frac{AUC}{1-AUC} = WR$.

### 2.3 Transitivity and Non-transitivity.

Transitivity and non-transitivity we define as the long-term behavior of random variables. For two random variables $x$ and $y$, we define $x \prec_{AUC} y$ as $AUC = \Pr(x < y) > 1/2$. The relationship $\prec_{AUC}$ is transitive if $x \prec_{AUC} y \prec_{AUC} z$ implies $x \prec_{AUC} z$. We define a non-transitive loop of length 3 as $x \prec_{AUC} y \prec_{AUC} z \prec_{AUC} x$. Similarly, non-transitivity with respect to the win ratio is defined when $x \prec_{WR} y \prec_{WR} z \prec_{WR} x$, where $x \prec_{WR} y$ means that the win ratio comparing $y$ to $x$ is greater than 1. We always assume that random variables $x, y$ and $z$ are independent.

3. Methods

**3.1 Existence of non-transitive loops formed by AUCs and win ratios of three independent random variables.**

*Statement 1. Existence.*

WR and AUC can form non-transitive loops of length 3.

*Proof.*

In the Appendix we show that the WR and AUC are monotonic transformations of each other. This implies that proving the non-transitivity of one is sufficient to guarantee the non-transitivity of the other.

We prove the existence of non-transitive loops of AUCs and WRs by example:

*Example 1. Non-transitive loop formed by discreet random variables.*

If $T_x, T_y$ and $T_z$ follow multinomial distributions with probability functions in Table 1, then $WR_{xy}>1.0$, $WR_{yz}>1.0$ and $WR_{zx}>1.0$. Indeed,

$Pr(x < y) = .4(.7 + .3) + .6 * .3 = .58, Pr(y < z) = .7, Pr(z < x) = .6$ and $WR_{xy} = \frac{.58}{.42} = 1.38$, $WR_{yz} = \frac{.6}{.4} = 1.5, WR_{zx} = \frac{.7}{.3} \approx 2.33$.

| r.v. | Pr | | | | |
|---|---|---|---|---|---|
| | 1 | 2 | 3 | 4 | 5 |
| x | .4 | 0 | 0 | .6 | 0 |
| y | 0 | .7 | 0 | 0 | .3 |
| z | 0 | 0 | 1 | 0 | 0 |

*Table 1. Three discreet non-transitive random variables.*

*Example 2. Continuous case.*

We use the following relationship. $Pr(x < y) = \int_{-\infty}^{+\infty} F(y) dG(y) = \int_{-\infty}^{+\infty} F(y) g(y) dy = \int_{-\infty}^{+\infty} (1 - G(x)) dF(x) = \int_{-\infty}^{+\infty} (1 - G(x)) f(x) dx$, where $x$ and $y$ are any independent random variables such that $x \sim F(\cdot), y \sim G(\cdot)$.

If $T_1 \sim \mathcal{X}_{df=1}, T_2 \sim N\left(\mu = \frac{1.175}{2}, \delta = 0.1\right)$ and $T_3 \sim 1.175 - \mathcal{X}_1$. Then, $WR_{12} > 1.0$, $WR_{23} > 1.0$ and $WR_{31} > 1.0$.

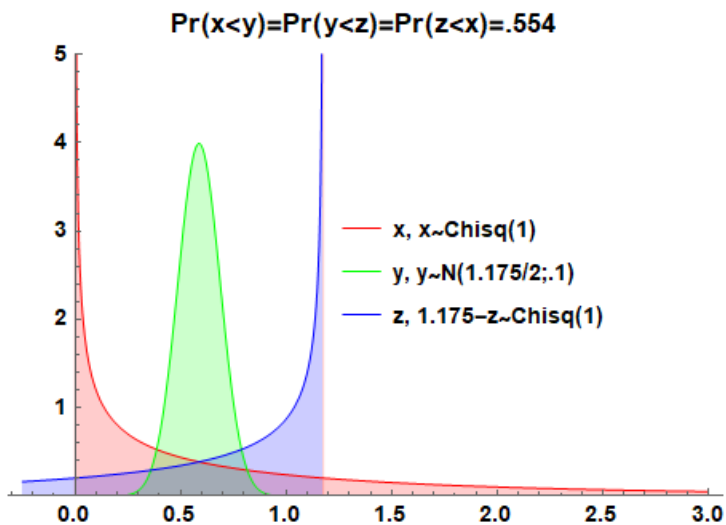

$\mu_x = 1.0; \mu_y = 0.5875; \mu_z = 1.175 - 1 = 0.175$

Examples 1 and 2 prove the existence of three random variables that form non-transitive loops with respect to AUC and consequently for the WR statistic.

∎

*Note 1*

The discrete distributions in example 1 are similar to the example from the Steinhaus paper on non-transitivity (Steinhaus and Trybula 1959).

*Note 2*

More examples of non-transitive loops are presented in the Appendix, including the famous example of Efron dice that form a non-transitive loop of length 4 (Gardner 1970).

**3.2 Sufficient conditions of transitivity of three independent random variables.**

*Lemma 1.*

*Three Binomial random variables taking values 0 and 1 are always transitive.*

If $x \sim Bin(p_x), y \sim Bin(p_y)$, then $\Pr(x < y) \geq .5$ iff $p_x < p_y$.

*Proof.*

Let's use the AUC as an example. $Pr(x < y) + .5Pr(y = x) = Pr(0,1) + .5(Pr(0,0) + Pr(1,1)) = (1 - p_x)p_y + .5\left((1 - p_x)(1 - p_y) + p_x p_y\right) = \frac{1 + p_y - p_x}{2}$. Therefore $AUC_{x<y} > .5$ iff $p_x < p_y$.

∎

*Lemma 2*

Three Normally distributed random variables are always transitive.

If $x \sim N(\mu_x, \sigma)$, $y \sim N(\mu_y, \tau)$, then $\Pr(x < y) \geq .5$ iff $\mu_x < \mu_y$.

*Proof.*

Indeed, $Pr(x < y) = Pr(x - y < 0)$, $x - y \sim N(\mu_x - \mu_y, \sqrt{\sigma^2 + \tau^2})$, so $Pr(x - y < 0) = Pr\left(z < \frac{\mu_y - \mu_x}{\sqrt{\sigma^2 + \tau^2}}\right) = \Phi\left(\frac{\mu_y - \mu_x}{\sqrt{\sigma^2 + \tau^2}}\right)$. Therefore $Pr(X < Y) \geq 1/2$ iff $\mu_x < \mu_y$. Because the comparison of means is always transitive, we conclude that three Normally distributed random variables are always transitive.

∎

*Lemma 3* Three Log-normally distributed random variables are always transitive.

*Proof*

$Pr(x < y)$ is invariant to monotonic transformation. Therefore $Pr(x < y) = Pr(e^x < e^y)$ and we can use Lemma 2 to infer transitivity.

∎

The results for normal and lognormal distribution functions can be generalized to any symmetric probability density function.

*Lemma 4.* Three random variables with symmetric pdfs are always transitive.

In other words, $Pr(x - y < 0) > \frac{1}{2}$ if and only if $M_x < M_y$, where $x$ and $y$ are independent random variables with symmetric probability density functions with points of symmetry $M_x \in \mathbb{R}$ and $M_y \in \mathbb{R}$ correspondingly.

*Proof*

First let's prove that if $Pr(x - y < 0) > \frac{1}{2} \implies M_x < M_y$. We note that random variables $x$ and $2M_x - x$ follow the same distribution. Similarly, random variables $y$ and $2M_y - y$ follow the same distribution. Then we can write: $\frac{1}{2} < Pr(x - y < 0) = Pr(x < y) = Pr(2M_x - x < 2M_y - y) = Pr(y - x < 2(M_y - M_x)) = Pr(x - y > 2(M_x - M_y)) = 1 - Pr(x - y < 2(M_x - M_y))$. In other words, $1 - Pr(x - y < 2(M_x - M_y)) > \frac{1}{2}$ or equivalently, $Pr(x - y < 2(M_x - M_y)) < \frac{1}{2}$. The two inequalities $Pr(x - y < 0) > \frac{1}{2}$ and $Pr(x - y < 2(M_x - M_y)) < \frac{1}{2}$, imply that $2(M_x - M_y) < 0$.

Next, we prove that $M_x < M_y \implies Pr(x - y < 0) > \frac{1}{2}$. We will do this by contradiction. Define $A = Pr(x - y < 0)$. Then $1 - A = 1 - Pr(x - y < 0)$. As before, we have that $1 - Pr(x - y < 0) = 1 - Pr(x < y) = 1 - Pr(2M_x - x < M_y - y) = 1 - Pr(y - x < 2(M_y - M_x)) = Pr(x - y < 2(M_x - M_y))$. Suppose that $A < \frac{1}{2}$. Since cumulative probability is defined as the integral of a non-negative function, and $M_x < M_y \implies 2(M_x - M_y) < 0$, we have that $Pr(x - y < 2(M_x - M_y)) < Pr(x - y < 0)$. This gives us that $\frac{1}{2} \leq 1 - A = Pr(x - y < 2(M_x - M_y)) < Pr(x - y < 0) = A < \frac{1}{2}$, i.e. $\frac{1}{2} \leq 1 - A < A < \frac{1}{2}$, which is a contradiction.

∎

When we completed this manuscript, we became aware of (Lebedev 2019) proof of this lemma.

Normal, Uniform, Student's t, Logistic, Cauchy and some other distributions have symmetric density functions and are therefore always transitive among each other. However, they can still form non-transitive relationships but only with random variables with asymmetric pdfs such as Chi-squared, exponential beta etc.

Trybula (Trybuła 1961) in Theorem 2 proved one more sufficient condition for transitivity.

*Lemma 5 (Trybuła 1961). Random variables with the same pdfs up to a shift are transitive.*

Three independent random variables that have common distribution functions up to a shift are always transitive. I.e. three random variables $x, y$ and $z$ that are independent and $Pr(x < t) = F(t), Pr(y < t) = F(t - d), Pr(z < t) = F(t - e)$, then $Pr(x < y) > \frac{1}{2}$ and $Pr(y < z) > \frac{1}{2}$ implies $Pr(x < z) > \frac{1}{2}$. The proof is in (Trybuła 1961).

(Lebedev 2019) proved that random variables with polynomial cumulative distributions functions defined on an interval can form non-transitive loops of length 3 but only for polynomials of degree 4 and above.

The next question is whether any two AUC values can guarantee that they cannot be part of a non-transitive loop of length 3.

*Lemma 6 (Trybuła 1965). For any three independent random variables x,y, and z, if $AUC_{xy} \cdot AUC_{yz} \geq \frac{1}{2}$, then $x, y$ and $z$ are transitive (that is $AUC_{zx} \leq 1/2$).*

*Proof*

In the corollary to Theorem 1 (equation 12), Trybula proved that for any number of independent random variables $x_1 \ldots x_n$, the inequality $AUC_{x_1 x_2} \cdot AUC_{x_2 x_3} \cdot \ldots \cdot AUC_{x_n x_1} \leq \frac{1}{4}$ always holds. Furthermore, this is a sharp upper bound, i.e. there are sets of random variables for which $AUC_{x_1 x_2} \cdot AUC_{x_2 x_3} \cdot \ldots \cdot AUC_{x_n x_1} = \frac{1}{4}$. Thus, for three random variables this implies that if $AUC_{xy} \cdot AUC_{yz} \geq \frac{1}{2}$ then $AUC_{zx} \leq \frac{1}{2}$ for the inequality to hold.

*Corollary 1.* If $AUC_{xy} \geq \frac{1}{\sqrt{2}} \approx .70711$ and $AUC_{yz} \geq \frac{1}{\sqrt{2}}$, then $AUC_{zx} \leq \frac{1}{2}$.

*Corollary 2.* If $WR_{xy} \geq \frac{1}{\sqrt{2}-1} \approx 2.41$ and $WR_{yz} \geq \frac{1}{\sqrt{2}-1}$, then $WR_{zx} \leq 1$.

For the Win Ratio, Lemma 6 cannot be translated into a compact expression as we need to check $\left(1 + \frac{1}{WR_1}\right)\left(1 + \frac{1}{WR_2}\right) \geq \frac{1}{2}$.

Indeed, since the WR and AUC are monotonic transformations of each other: $\left(1 + \frac{1}{WR_1}\right)\left(1 + \frac{1}{WR_2}\right) \geq \frac{1}{\sqrt{2}}$

### 3.3. Necessary conditions for non-transitive loops of length 3.

*Lemma 7 (Savage Jr 1994, Богданов 2010, Komisarski 2021).* $\sup_{S_3} \min (AUC) = \frac{\sqrt{5}-1}{2}$.

If three r.v. $x, y$ and $z$ form a non-transitive loop, then at least one of their AUCs is less than or equal to $\frac{\sqrt{5}-1}{2} \approx 0.618$. In other words, $\sup_{S_3} \min (AUC) = \frac{\sqrt{5}-1}{2}$, where supremum is taken over the space $S_3$: the space of all non-transitive loops of length 3 and minimum is calculated within the loop.

*Proof.*

The discrete case follows directly from Theorem 1 in Savage "Paradox of nontransitive dice", which states:

Savage Theorem 1: "Suppose numbers $1, 2, \ldots, 3n$ are arranged on the faces of 3 [non-transitive] n-sided dice. Then at least one of the probabilities is less than $\frac{\sqrt{5}-1}{2}$."

The continuous case follows from Lemma 2 in (Trybuła 1961) which states that the AUC of continuous random variables can be approximated with the AUC of discrete random variable as closely as possible.

∎

Corollary: Thus, a non-transitive loop of length 3 cannot have all AUCs greater than $\frac{\sqrt{5}-1}{2} \approx 0.618$.

Bogdanov (Богданов 2010) and Komisarski (Komisarski 2021) showed that $\sup_{S_n} \min (AUC) = 1 - \frac{1}{4\cos^2 \frac{\pi}{n+2}}$, where $S_n$ is the space of sets of AUCs for non-transitive loops of $n$ independent discrete random variables. Interestingly, this implies that the minimum of the AUCs within the loop goes up with the length of the loop. For n=3, $\sup_{S_3} \min (AUC) = 1 - \frac{1}{4\cos^2 \frac{\pi}{3+2}} = \frac{\sqrt{5}-1}{2}$. Note that $\frac{\sqrt{5}-1}{2}$ is the Golden Ratio (GR). As $n$ goes to infinity, the supremum reaches a maximum of ¾, and therefore a non-transitive loop of any length cannot have all $AUC$ values $> .75$.

Statement 1 and corollaries to Lemma 6 provide thresholds for the AUC and win ratio summarized in Table 2.

| WR range | AUC range | Strength | Explanation |
|---|---|---|---|
| $\left[1.00; \dfrac{\sqrt{5}-1}{3-\sqrt{5}}\right]$ $\approx [1.00; 1.62]$ | $\left[.50; \dfrac{\sqrt{5}-1}{2}\right]$ $\approx [.50; .62]$ | Weak | Necessary condition for non-transitivity is satisfied with any third variable (Lemma 7). |
| $\left(\dfrac{\sqrt{5}-1}{3-\sqrt{5}}; \dfrac{1}{\sqrt{2}-1}\right]$ $\approx (1.62; 2.41]$ | $\left(\dfrac{\sqrt{5}-1}{2}; \dfrac{1}{\sqrt{2}}\right]$ $\approx (.62; .71]$ | Moderate | Necessary condition of non-transitivity is not satisfied but it is not high enough for transitivity with any third variable (Corollary 1 of Lemma 6). |
| $\left(\dfrac{1}{\sqrt{2}-1}; 3.00\right]$ $\approx (2.41; 3.00]$ | $\left(\dfrac{1}{\sqrt{2}}; \dfrac{3}{4}\right]$ $\approx (.71; .75]$ | Medium | Adding a third variable that has an AUC of at least the same strength guarantees transitivity of the corresponding three random variables (Corollary 1 of Lemma 6). |
| $(3.00; +\infty)$ | $(.75; 1.00]$ | Strong | Transitivity guaranteed when other AUCs have at least the same strength (Lemma 7). |

*Table 2. Thresholds of AUC and WR with respect to their ability to be part of transitive loops.*

We can fully define the space of non-transitive loops of AUCs of length 3.

*Lemma 8 (Trybuła 1961).*

The space of AUCs generated by triplets of any independent random variables $X', Y'$ and $Z'$ with non-overlapping values such that $\Pr(X' = c) = \Pr(Y' = c) = \Pr(Z' = c) = 0$ for any constant $c \in \mathbb{R}$, is bounded from above by the AUCs generated by random variables from a family of distributions $\Sigma$. Here $\Sigma$ is defined as a family of distributions of three independent multinomial random variables $X, Y$ and $Z$ such that $Pr(X = x_1) + Pr(X = x_2) = p_x + q_x = 1$, $Pr(Y = y_1) + Pr(Y = y_2) = p_y + q_y = 1, Pr(Z = z) = 1$ and $x_1 < y_1 < z < x_2 < y_2$. Two examples of distributions from Σ are provided in Figure 2. (Trybuła 1961) prove Lemma 8.

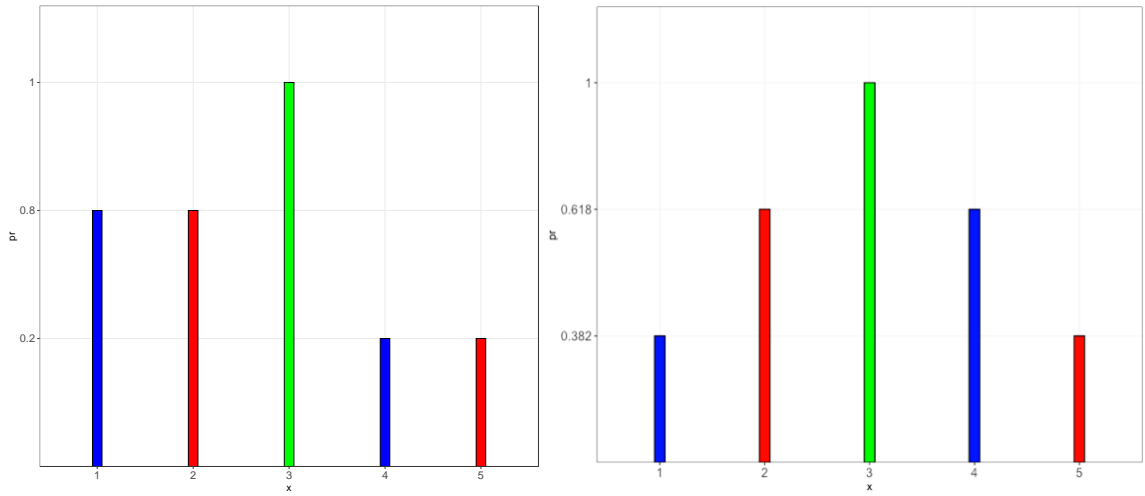

*Figure 2. Multinomial distributions of three independent variables (x blue, y red and z in blue) from family of distributions Σ. Left: transitive example, right non-transitive example.*

*Lemma 9. Space of three non-transitive AUCs.*

*The space of AUCs $S_{3nT} = A, B, C$ formed by AUCs of three non-transitive independent random variables is fully defined up to their permutation as:*

$$\begin{cases} A = 1 - BC \\ \frac{1}{2} \leq B \leq 1, \\ \frac{1}{2} \leq C \leq \frac{1}{2B} \end{cases}$$

*Where $A = AUC_{xy} = Pr(x < y), B = AUC_{yz} = Pr(y < z), C = AUC_{zx} = Pr(z < x)$.*

*Proof*

Step 1. Upper boundary of space $S_{3nT}$.

We will use Lemma 8, that for each triplet of independent random variables $X', Y'$ and $Z'$, there exists a triplet $X, Y, Z$ from class Σ defined above such that at least one of its permutations satisfies the set of inequalities $\Pr(X < Y) \geq \Pr(X' < Y'), \Pr(Y < Z) \geq \Pr(Y' < Z')$, and $\Pr(Z < X) \geq \Pr(Z' < X')$. This result implies that to define an upper AUC boundary for $S_{3nT}$, it is enough to define the subspace of non-transitive triplets from Σ.

If $X, Y, Z \in \Sigma$ then:

$$A = \Pr(X < Y) = p_x p_y + p_x(1 - p_y) + (1 - p_x)(1 - p_y) = 1 - BC \geq 1/2 \Rightarrow C \leq \frac{1}{2B}$$

$$B = \Pr(Y < Z) = p_y \implies \frac{1}{2} \leq B \leq 1$$

$$C = \Pr(Z < X) = 1 - p_x \implies \frac{1}{2} \leq C \leq \frac{1}{2B}$$

Step 2. Lower boundary of space $S_{3nT}$.

To define the lower boundary we use Theorem 1 from (Trybuła 1961) which we reproduce below:

Theorem 1 (Trybuła 1961): For any triplet of AUCs ($A, B$ and $C$) there exist three independent random variables such that $\Pr(X = x) = \Pr(Y = y) = \Pr(Z = z) = 0$ that generate these AUCs if and only if their AUCs (A, B and C) satisfy these inequalities:

$$1 - \alpha(1 - A, 1 - B) \leq C \leq \alpha(A, B), \tag{1}$$

where

$$\alpha(A, B) = \begin{cases} max\left(\frac{1-A}{B}, \frac{1-B}{A}, 1 - AB\right), & if\ A + B > 1 \\ 1, & if\ A + B \leq 1 \end{cases}$$

This theorem states the necessary and sufficient conditions that define the space $S_3$ formed by the AUCs of any three independent random variables. We want to define a subspace $S_{3nT} \subset S_3$ formed by non-transitive loops. In other words we restrict $A, B, C$ such that $A \geq \frac{1}{2}, B \geq \frac{1}{2}, C \geq \frac{1}{2}$. Which implies that $A + B > 1$. In this case the left hand side of the inequality (1) reduces to trivial 0 because $\alpha(1 - A, 1 - B) = 1$ and hence the left hand side in general is bounded by 0 and can be replaced for non-transitive case with 1/2.

∎

We plot the space $S_3$ and $S_{3nT}$ of non-transitive triplets in Figure 3 and similar plots for the WR in the Appendix.

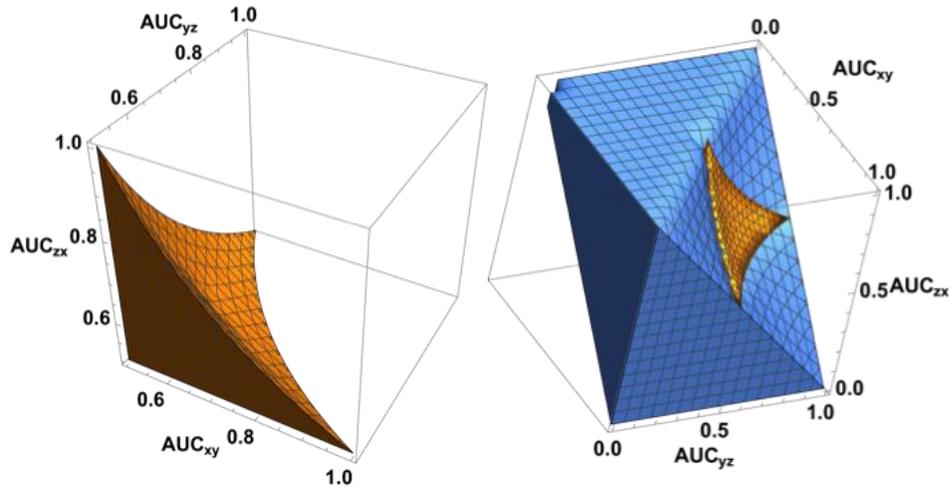

*Figure 3. Space $S_{3nT}$ of non-transitive triplets of AUC (left) and space $S_3$ (right) of all possible triplets of AUC comparing any three independent random variables.*

For medical applications, it is important to know if there is an AUC high enough that it rules out non-transitivity. In other words, it is important to find $\sup_{S_n} \mathbf{max}(AUC)$, to guarantee that there can be no instances of intransitivity in a dataset. This question can be answered for $n = 3$ random variables.

*Lemma 10.* $\sup_{S_3} \mathbf{max}(AUC) = 1.0$

*Proof*

It is enough to show that $(.5, 1.0, .5) \in S_{3nT}$.

∎

Lemma 10 implies that a single AUC as high as 1.0 can be part of non-transitive loop.

4. **Practical Examples.**

We use Project Data Sphere (Bertagnolli et al. 2017, Green et al. 2015) to download person-level data for three randomized controlled trials, each with three or more treatment arms. We use the win ratio to compare the time to all-cause death. Further details on how we deal with censoring when computing win ratios is provided in the Appendix.

*Example 1.*

Trial 1 N=283 compares two treatments to a placebo. The win ratios and corresponding AUCs for this trial are presented in Table 3.

|       | Win Ratio | AUC |
|-------|-----------|-----|
| $A > B$ | 1.39332 | 0.58217037 |
| $A > C$ | 1.89 | 0.65397924 |
| $A > D$ | 2.341538 | 0.70073661 |

*Table 3. Win ratios and AUCs comparing time to death in randomized controlled trial NCT00785291 N=283 with three treatment arms A (N=51), B (N=8) and C (N=224).*

This trial has transitive win ratios. The time to death observed for participants from arm A is longer than for arm B, which is longer than for time C, when measured by win ratio (WR of 1.39, 1.89 and 2.34 correspondingly). The win ratio of 1.39 is weak and can satisfy the necessary condition for non-transitivity when compared to other treatments. Therefore, we should beware of chains of comparisons that include treatments A and B. Such chains can be non-transitive as they include this weak comparison. The comparison of time to death between B and C is stronger (moderate according to our classification in Table 3) with a win ratio of 1.89. The strongest comparison is for time to death of participants from treatment groups A versus C (WR of .70), but it still remains only moderate.

*Example 2.*

Trial 2 N=3,871 compares two treatments to a placebo. The win ratios and corresponding AUCs for this trial are presented in Table 4.

|       | Win Ratio | AUC |
|-------|-----------|-----|
| $A > B$ | 1.03971 | 0.50973423 |
| $A > C$ | 1.22148 | 0.54984965 |
| $A > D$ | 1.36922 | 0.57792016 |
| $B > C$ | 1.317021 | 0.56841134 |
| $B > D$ | 1.199293 | 0.54530842 |
| $C > D$ | 1.097568 | 0.52325741 |

*Table 4. Win ratios and AUCs comparing time to death in a randomized controlled trial NCT00041119 N=3,871 with three treatment arms A (N=1,142), B (N=789), C (N=1,151) and D (N=789).*

This trial has transitive win ratios. The time to death observed for participants from arm A is longer than for arm B, which is longer than for time C, which is longer for arm D when measured by win ratio (WR of 1.04, 1.22 and 1.1 correspondingly). This comparison is transitive as the time to death observed for participants from arm A is longer than for arm D with a win ratio of 1.37. However, the strength of stochastic comparison between all these treatments ranges from weak to moderate. Therefore, we should beware of chains of comparisons that include these treatments. Some of these comparisons can satisfy the necessary condition for non-transitivity when compared to other treatments.

*Example 3.*

Trial 3 N=25,871 compares two treatments to a placebo. The win ratios and corresponding AUCs for this trial are presented in Table 5.

| | All-cause mortality | | | Myocardial infarction | |
|---|---|---|---|---|---|
| | Win Ratio | AUC | | Win Ratio | AUC |
| $A > B$ | 1.0005 | 0.5001 | $B > A$ | 1.043 | 0.5106 |
| $A > C$ | 1.0111 | 0.5028 | $C > A$ | 1.410 | 0.5850 |
| $D > A$ | 1.0230 | 0.5057 | $D > A$ | 1.445 | 0.5909 |
| $B > C$ | 1.0105 | 0.5026 | $C > B$ | 1.351 | 0.5747 |
| $B > D$ | 1.0238 | 0.5059 | $D > B$ | 1.385 | 0.5806 |
| $C > D$ | 1.0346 | 0.5085 | $C > D$ | 1.024 | 0.5060 |

*Table 5. Win ratios and AUCs comparing time to death (left) and myocardial infarction (right) in randomized controlled trial NCT01169259 N=25,871 with three treatment arms A (N=6,474), B (N=6,464), C (N=6,470), and D (N=6,463). WRs (left) based on all-cause mortality form three non-transitive loops: $A >_{WR} B >_{WR} C >_{WR} D >_{WR} A$ ; $A >_{WR} C >_{WR} D >_{WR} A$; $B >_{WR} C >_{WR} D >_{WR} A$. WRs (right) based on time to first myocardial infarction is transitive.*

This trial produces very weak and non-transitive win ratios when comparing times to death. The time to death observed for participants from arm A is longer than for arm B, which is longer than for time C, which is longer than arm D, which is longer than arm A (WR of 1.0005, 1.0105, 1.0346, and 1.0230 correspondingly). The strength of stochastic comparisons, based on time to all-cause death, is very weak (all WRs<1.04). The trial was not designed to measure all-cause death. A longer follow-up could alter the win ratios. Stronger comparisons are observed for myocardial infarction which was one of the secondary endpoints of the trial. Table 5 (righthand side) summarizes win ratios for myocardial infarction. WRs for this outcome are transitive.

5.  **Discussion and Outlook**

In this work we demonstrate that the win ratio, a novel statistic currently used to evaluate the efficacy of treatment in several large randomized controlled trials, can form paradoxical non-transitive loops when comparing three or more treatment arms. We study the conditions that guarantee transitivity in the case of three independent random variables. We establish that even in cases in which the WR and AUC have maximal strength, they can be a part of non-transitive loops of length 3. We establish that three (or more) random variables with symmetric pdfs cannot form non-transitive loops. We also prove that if a product of two AUCs is greater than $\frac{1}{2}$, then they cannot be part of a non-transitive loop of length 3. The necessary condition for non-transitivity states that a non-transitive loop of length $n$ contains at least one win ratio that is less than or equal to $4\cos^2\frac{\pi}{n+2}$, and for the AUC this condition is $AUC \leq 1 - \frac{1}{4\cos^2\frac{\pi}{n+2}}$. In special cases when n=3 the AUC value for this boundary becomes the golden ratio, as $n \to \infty$ it goes to 3.0 for WR and ¾ for AUC. We formulate thresholds on the AUC and WR that provide guarantees of various strength of transitivity. In other words, we define the boundaries within which transitivity does not hold.

**5.1 Non-transitivity reports across different fields of research.**

In this work, we study the non-transitive long-run behavior of the win ratio. We define non-transitivity on the level of distribution functions of random variables using a similar definition as Steinhaus, Trybula and Savage Jr, Efron, Bogdanov, Komisarski and others. Non-transitivity is commonly reported in sports (Temesi, Szádoczki, and Bozóki 2023), particularly in chess (Poddiakov 2022, Sanjaya, Wang, and Yang 2022). In econometrics it is studied in game theory when decisions are made under uncertainty and risk. In quantum mechanics, outcomes of classical measurements are intrinsically probabilistic and non-transitivity is also possible, as reported in (Stohler and Fischbach 2005, Ramzan, Khan, and Khan 2010). (Gorbunova and Lebedev 2022) study the effect of non-transitivity in applications to Bayesian queueing models. In medical literature non-transitivity is reported in network meta-analysis and is attributed to heterogeneity of compared populations (Cipriani et al. 2013). Several authors (Buyse 2010, Rauch et al. 2018, Brunner 2020) note the possibility of non-transitive behavior of the win ratio. (Johns et al. 2023) notes that the field is split over whether non-transitivity of the win ratio is a benefit or a disadvantage. (Lumley and Gillen 2016) state that non-transitivity in the context of hypothesis testing is a theoretical flaw of employed test statistics such as AUC, arguing that if the lack of transitivity in a statistic can pose difficulties in the specific context, then non-transitive statistics should be avoided. (Brunner et al. 2021, Brunner, Bathke, and Konietschke 2018, Thangavelu and Brunner 2007) propose potential solutions to address the non-transitivity problem in statistical tests like Kruskal-Wallis and van Elteren, among others. On the contrary, (Poddiakov 2023, Poddiakov 2018) argue that non-transitivity is a natural phenomenon with practical utility in various fields including biology. (Poddiakov and Lebedev 2023) also outline a method for constructing non-transitive sets. Examples of useful applications of non-transitivity in biology can be found in (Lehtinen et al. 2022) and (Liao et al. 2020). Liao references numerous instances of non-transitivity in bacteriology where it is hypothesized that non-transitivity is essential for avoiding dominance of one species and maintaining the coexistence of diverse bacteria strains. In the present study we provide an example of a clinical trial, where the win ratio is non-transitive and argue that non-transitivity is a sign of the weakness of the treatments' efficacy that is undetectable by statistics that can only be transitive.

**5.2 Non-transitivity as an indicator of the weakness of stochastic comparisons.**

We use this paper to make a broader argument for the importance of reporting the strength of stochastic comparisons. Interestingly, comparing the summary statistics calculated among patients within the same group is always transitive: the restricted mean or median survival time and the difference in group means in clinical trials with soft endpoints (decrease in blood pressure, LDL cholesterol etc.). (Lumley and Gillen 2016) provide a rigorous proof demonstrating that only tests based on univariate within-group statistics can guarantee transitivity. Thus, any statistics based on head-to-head comparisons (i.e. those that combine data from pairs of participants from two different groups) can result in this non-transitive effect. We claim that the possibility of non-transitivity is essential information for evaluating the benefit (or lack) of one treatment over the other. For example, let's consider two clinical trials with three arms comparing six different treatments. Suppose that both trials have the same three different median survival times for each treatment, and that in one trial the win ratios for time to all-cause death form a non-transitive loop, while in the other trial the win ratios are transitive and consistent with the directionality of comparisons of median survival times. In this case, we should interpret the non-transitivity in the first trial as an absence of a clear winner, while in the transitive trial the winning treatment is unambiguous. Metrics that can be transitive evaluate the strength of stochastic relationships between two random variables. These metrics add valuable and complementary information about the strength of stochastic comparisons which so far has not been explicitly studied in medical research. In practice, we suggest expanding the set of statistics that are reported in clinical trials by including statistics such as win ratios and AUCs. We also suggest checking whether pairwise comparisons of two treatments exceeds the natural thresholds for win ratios of 1.62, 2.41 and 3.0 (for AUC of .62, .71 and ¾). Further research is needed to define what other estimates of strength of stochastic relationships can complement traditional statistics.

**5.3 Limitations**

A limitation of this study is the weakness of the win ratios observed in the non-transitive example in a real-life clinical trial. However, our theoretical results show that having strong win ratios does not guarantee transitivity and it is possible that in another clinical trial, stronger win ratios can form a non-transitive loop, even in cases where one of the AUCs is close to 1. We also use a simplified definition of the win ratio that considers only one outcome and ignores the hierarchy of other outcomes. However, the win ratio is based on head-to-head comparisons, and it is likely that non-transitivity is possible for win ratios with hierarchical outcomes as well.

**5.4 Future Directions**

Much more work is required to study the strength of stochastic relationships in medical research. The first question that should be addressed is which existing statistics measure the strength of stochastic relationships. (Lumley and Gillen 2016) proves that confining the search to univariate within-group statistics is the only way to ensure transitivity. (Gillen and Emerson 2007) show that (weighted) log-rank tests can be non-transitive. Given the relationship of the win ratio and hazard ratio, the latter can be non-transitive. Finally, we want to know how the strength of stochastic relationships can benefit clinical trials and the medical community in general. While non-transitivity of efficacy measures in randomized double-blinded controlled trials is "induced" by distribution functions of the outcome in trial arms and is mostly not affected by confounding, non-transitivity in observational studies is affected by confounding and is not understood well. Another interesting direction is non-transitivity of repeated measures. When random variables are correlated, their non-transitive behavior is not described well.

6. **Conclusions**

Transitivity is an assumed property in contemporary medical research. However, this study demonstrates that transitivity cannot be taken for granted as demonstrated by the practical example in one of the publicly available large randomized controlled trials. In this paper we study the non-transitivity of the win ratio and AUC. We summarize published results and add new sufficient conditions of transitivity. We also study the necessary conditions of non-transitivity and show that the space of non-transitive loops of length three comes infinitely close to 1.0 for the AUC and to infinity for the win ratio, implying that even close to perfect comparisons can be part of a non-transitive loop of length three. We argue that non-transitivity is a lack of strength of the stochastic relationship between two groups. Therefore, the possibility of non-transitivity is not a deficiency of the AUC and win ratio but rather a valuable property that shouldn't be ignored. We note that many statistics including some that are used to report efficacy in randomized controlled trials are not designed to "detect" non-transitivity, while metrics that explicitly measure the strength of stochastic order relationships (such as the win ratio) can detect non-transitivity. All of this points to the fact that measures of the strength of stochastic relationships can provide valuable information on the efficacy of treatment in clinical trials.

The results of this paper demonstrate that the win ratio, as a metric of the strength of stochastic relationships, can detect non-transitive relationships in a randomized controlled trial, while a comparison of means cannot. We argue that the strength of stochastic relationships should be explicitly assessed and reported in order to fully understand all aspects of treatment efficacy.

**Appendix**

*Statement A1* WR, Mann-Whitney statistic and AUC are strictly monotonic transformation of each other.

*Proof.*

Indeed:

$$WR = \frac{\sum I[T_{1i}>T_{2j}]}{\sum I[T_{1i}<T_{2j}]} = \frac{p}{q}.$$

Bamber and MacNeil [] showed that Area under the Receiver Operating Characteristics Curve (AUC or AUCROC) comparing random variable $T_1$ in group 1 to $T_2$ in group 2 can be written as $AUC_{12} = \Pr(T_1 > T_2)$. Therefore we can write $AUC = \frac{\sum I[T_{1i}>T_{2j}]}{\#\ comparisons} = \frac{\sum I[T_{1i}>T_{2j}]}{\sum I[T_{1i}>T_{2j}]+\sum I[T_{1i}<T_{2j}]} = \frac{p}{p+q}$ or $\frac{1}{AUC} = \frac{p+q}{p} = 1 + \frac{1}{WR} = \frac{WR+1}{WR}$ or $AUC = \frac{WR}{WR+1}$ or $\frac{1}{AUC} - 1 = \frac{1-AUC}{AUC} = \frac{1}{WR}$ or $\frac{AUC}{1-AUC} = WR$, where we denoted as $p = \sum I[T_{1i} > T_{2j}]$ and $q = \sum I[T_{1i} < T_{2j}]$.

It is easy to show that $AUC = \frac{WR}{1+WR}$ and $WR = \frac{AUC}{1-AUC}$. WR and AUC are monotonic transformations of each other by calculating the derivative of $f(x) = \frac{x}{1+x}$:

$$f'(x) = \left(\frac{x}{1+x}\right)' = \frac{1+x-x}{(1+x)^2} = \frac{1}{(1+x)^2} > 0 \text{ for all real values of x.}$$

Thus, we conclude that the AUC and WR are monotonic increasing transformations of each other. Similarly, we note that Mann-Whitney statistics are equivalent to AUC.

∎

*Corollary. To test WR we can use tests developed for AUC using U-statistics theory.*

*Statement A2*

If $T_1, T_2, T_3$ form non-transitive loop based on one of the three measures (WR, AUC or MW), then $T_1, T_2, T_3$ form non-transitive loop based on the other two, i.e.

$WR_{12}$>1.0, $WR_{23}$>1.0 and $WR_{31} > 1.0 \Leftrightarrow AUC_{12}$>0.5, $AUC_{23}$>0.5 and $AUC_{31}$>0.5 $\Leftrightarrow$ $MW_{12}, MW_{23}$ and $MW_{31}$.

*Proof.*

Lets define non-transitivity in AUC as:

$$AUC_{12} = p_{12} = \frac{\sum I[T_{1i} > T_{2j}]}{n_{12}} > .5$$

$$AUC_{23} = p_{23} = \frac{\sum I[T_{2j} > T_{3k}]}{n_{23}} > .5$$

$$AUC_{32} = p_{32} = \frac{\sum I[T_{3k} > T_{2i}]}{n_{23}} > .5$$

∎

*Examples of random variables that form non-transitive loops.*

*Example A1. Efron Dice.*

.

*Example A2.*

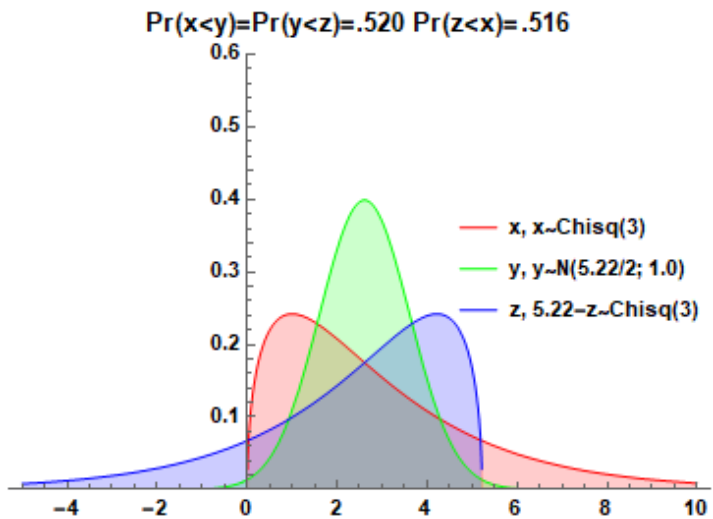

Figure A2. PDFs of three non-transitive continuous random variables.

$\mu_x = 3.0;\ \mu_y = 2.61;\ \mu_z = 5.22 - 3 = 2.22$

*Example A3.*

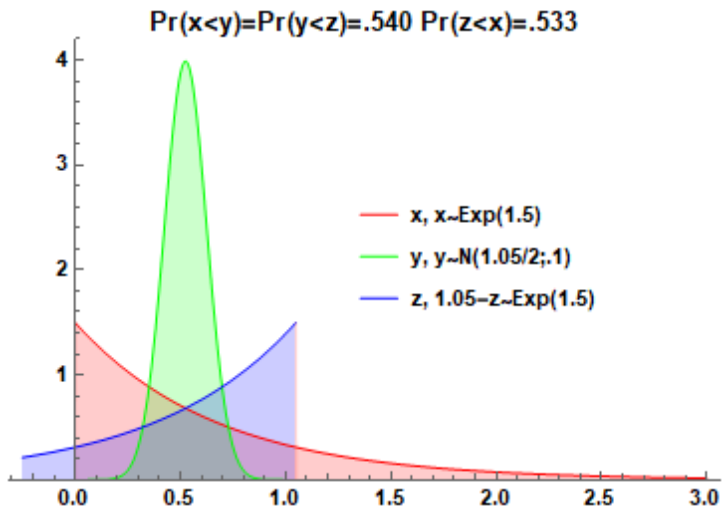

Figure A3. PDFs of three non-transitive continuous random variables.

$\mu_x = 0.667$; $\mu_y = 0.525$; $\mu_z = 1.0 - .667 = 0.333$

*Example A4.*

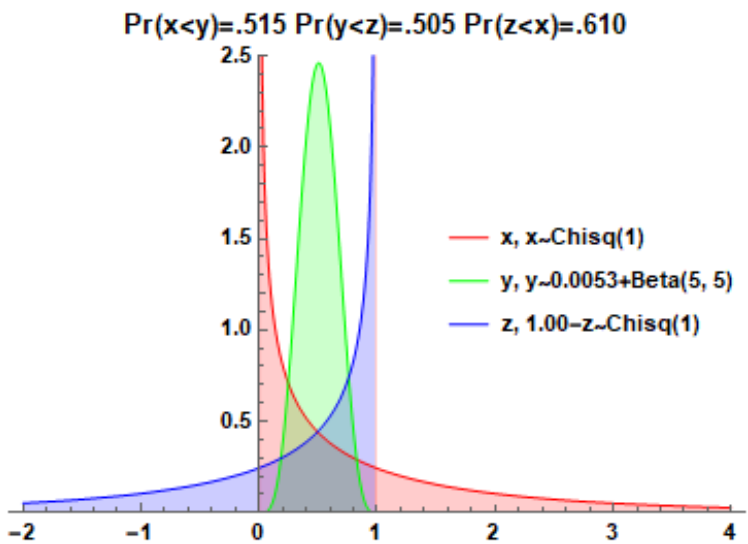

Figure A4. PDFs of three non-transitive continuous random variables.

$\mu_x = 1.0$; $\mu_y = .0053 + \dfrac{5}{5+5} = .5053$; $\mu_z = 0$

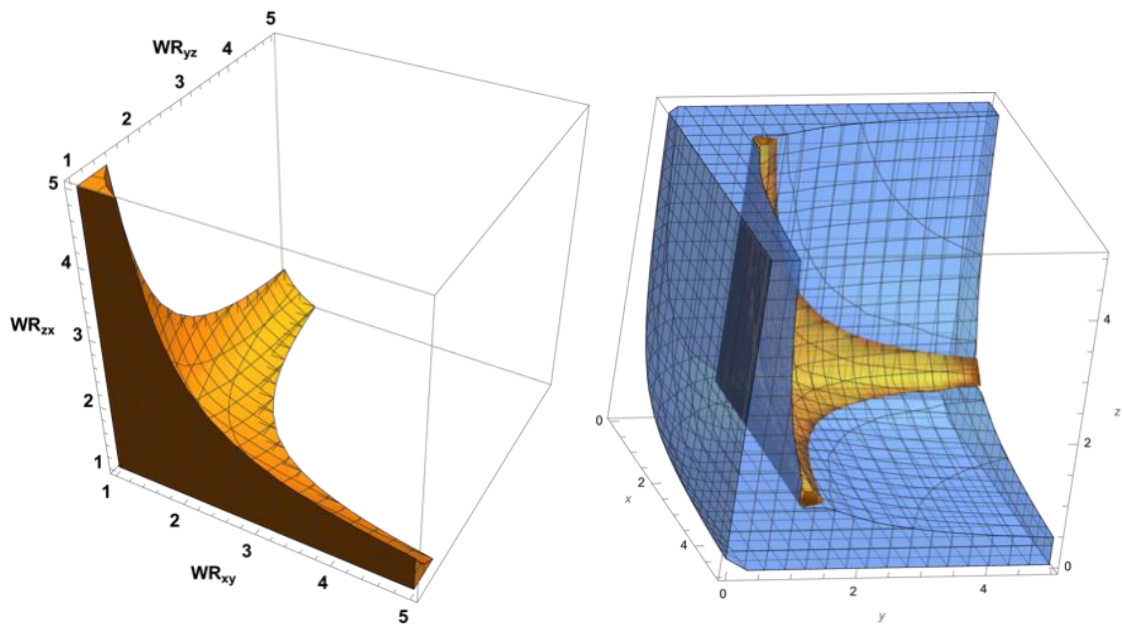

*Figure A1. Space of triplets of WRs formed by three independent random variables. A. non-transitive subspace; B. Blue: transitive subspace; Yellow: non-transitive subspace.*

*Computing win ratios in the presence of censoring*

We used the definition of win ratio from (Finkelstein and Schoenfeld 1999) (Pocock et al. 2012). For the simplicity of presentation, in all these examples we did not take into consideration the hierarchy of events. The time to all-cause death was considered censored at the end of follow-up. When comparing time to myocardial infarction, only the time to first occurrence of myocardial infarction was considered and all other follow-up times were considered censored at the end of follow-up. Wins were defined in the following way: when comparing times to event or censoring for a given participant from treatment group A to a participant from treatment group B ($T_A$ and $T_B$ correspondingly), treatment A "wins" over treatment B if $T_A > T_B$ and participant from group B had an event; treatment B "wins" over treatment A if $T_B > T_A$ and participant from group A had an event.

**Acknowledgments**

We are grateful to Professors Bradley Efron and Yaakov Malinovsky for a helpful discussion of an early results of this article.

**Funding**

IAD received funding from the Caltech EAS Chair Scholar Award and NSF Graduate Research Fellowship. OVD received funding from the NIH NHLBI grant R21HL167173, 5K01HL135342 and Swiss Federal Institute of Technology (ETH).

**Disclosure Statement**

Olga Demler have received funding from Kowa Research Institute for work unrelated to the current study.

**References**

Bamber, Donald, and Barbara J McNeil. 1975. "The area above the ordinal dominance graph and the area below the receiver operating characteristic graph." *Journal of mathematical psychology* no. 12 (4):387-415.

Benatar, Michael, Joanne Wuu, Peter M Andersen, Nazem Atassi, William David, Merit Cudkowicz, and David Schoenfeld. 2018. "Randomized, double-blind, placebo-controlled trial of arimoclomol in rapidly progressive SOD1 ALS." *Neurology* no. 90 (7):e565-e574.

Bertagnolli, Monica M, Oliver Sartor, Bruce A Chabner, Mace L Rothenberg, Sean Khozin, Charles Hugh-Jones, David M Reese, and Martin J Murphy. 2017. "Advantages of a truly open-access data-sharing model." *The New England journal of medicine* no. 376 (12):1178-1181.

Brunner, Edgar. 2020. "Success-Odds: An improved Win-Ratio." *arXiv preprint arXiv:2002.09273*.

Brunner, Edgar, Arne C Bathke, and Frank Konietschke. 2018. *Rank and pseudo-rank procedures for independent observations in factorial designs*: Springer.

Brunner, Edgar, Frank Konietschke, Arne C Bathke, and Markus Pauly. 2021. "Ranks and Pseudo-ranks—Surprising Results of Certain Rank Tests in Unbalanced Designs." *International Statistical Review* no. 89 (2):349-366.

Buyse, Marc. 2010. "Generalized pairwise comparisons of prioritized outcomes in the two-sample problem." *Statistics in medicine* no. 29 (30):3245-3257.

Cipriani, Andrea, Julian PT Higgins, John R Geddes, and Georgia Salanti. 2013. "Conceptual and technical challenges in network meta-analysis." *Annals of internal medicine* no. 159 (2):130-137.

De Neve, Jan, and Thomas A Gerds. 2020. "On the interpretation of the hazard ratio in Cox regression." *Biometrical Journal* no. 62 (3):742-750.

Dong, Gaohong, David C Hoaglin, Junshan Qiu, Roland A Matsouaka, Yu-Wei Chang, Jiuzhou Wang, and Marc Vandemeulebroecke. 2019. "The win ratio: on interpretation and handling of ties." *Statistics in Biopharmaceutical Research*.

Ferreira, João Pedro, Pardeep S Jhund, Kévin Duarte, Brian L Claggett, Scott D Solomon, Stuart Pocock, Mark C Petrie, Faiez Zannad, and John JV McMurray. 2020. "Use of the win ratio in cardiovascular trials." *Heart failure* no. 8 (6):441-450.

Finkelstein, Dianne M, and David A Schoenfeld. 1999. "Combining mortality and longitudinal measures in clinical trials." *Statistics in medicine* no. 18 (11):1341-1354.

Gardner, Martin. 1970. "Paradox of nontransitive dice and elusive principle of indifference." *Scientific American* no. 223 (6):110-&.

Gillen, Daniel L, and Scott S Emerson. 2007. "Nontransitivity in a class of weighted logrank statistics under nonproportional hazards." *Statistics & probability letters* no. 77 (2):123-130.

Gorbunova, AV, and AV Lebedev. 2022. "Nontransitivity of tuples of random variables with polynomial density and its effects in Bayesian models." *Mathematics and Computers in Simulation* no. 202:181-192.

Green, Angela K, Katherine E Reeder-Hayes, Robert W Corty, Ethan Basch, Mathew I Milowsky, Stacie B Dusetzina, Antonia V Bennett, and William A Wood. 2015. "The project data sphere initiative: accelerating cancer research by sharing data." *The oncologist* no. 20 (5):464-e20.

Johns, Hannah, Bruce Campbell, Julie Bernhardt, and Leonid Churilov. 2023. "Generalised pairwise comparisons for trend: An extension to the win ratio and win odds for dose-response and prognostic variable analysis with arbitrary statements of outcome preference." *Statistical Methods in Medical Research* no. 32 (3):609-625.

Komisarski, Andrzej. 2021. "Nontransitive random variables and nontransitive dice." *The American Mathematical Monthly* no. 128 (5):423-434.

Lebedev, Alexey V. 2019. "The nontransitivity problem for three continuous random variables." *Automation and Remote Control* no. 80:1058-1068.

Lehtinen, Sonja, Nicholas J Croucher, François Blanquart, and Christophe Fraser. 2022. "Epidemiological dynamics of bacteriocin competition and antibiotic resistance." *Proceedings of the Royal Society B* no. 289 (1984):20221197.

Liao, Michael J, Arianna Miano, Chloe B Nguyen, Lin Chao, and Jeff Hasty. 2020. "Survival of the weakest in non-transitive asymmetric interactions among strains of E. coli." *Nature communications* no. 11 (1):6055.

Lumley, Thomas, and Daniel L Gillen. 2016. "Characterising transitive two-sample tests." *Statistics & Probability Letters* no. 109:118-123.

Maurer, Mathew S, Jeffrey H Schwartz, Balarama Gundapaneni, Perry M Elliott, Giampaolo Merlini, Marcia Waddington-Cruz, Arnt V Kristen, Martha Grogan, Ronald Witteles, and Thibaud Damy. 2018. "Tafamidis treatment for patients with transthyretin amyloid cardiomyopathy." *New England Journal of Medicine* no. 379 (11):1007-1016.

Novack, Victor, Jeremy R Beitler, Maayan Yitshak-Sade, B Taylor Thompson, David A Schoenfeld, Gordon Rubenfeld, Daniel Talmor, and Samuel M Brown. 2020. "Alive and Ventilator-Free: A Hierarchical, Composite Outcome for Clinical Trials in the Acute Respiratory Distress Syndrome." *Critical care medicine* no. 48 (2):158.

Pocock, Stuart J, Cono A Ariti, Timothy J Collier, and Duolao Wang. 2012. "The win ratio: a new approach to the analysis of composite endpoints in clinical trials based on clinical priorities." *European heart journal* no. 33 (2):176-182.

Poddiakov, Alexander. 2018. "Intransitive machines." *arXiv preprint arXiv:1809.03869*.

Poddiakov, Alexander. 2022. "Intransitively winning chess players positions." *arXiv preprint arXiv:2212.11069*.

Poddiakov, Alexander, and Alexey V Lebedev. 2023. "Intransitivity and meta-intransitivity: meta-dice, levers and other opportunities." *European journal of mathematics* no. 9 (2):27.

Poddiakov, Alexander N. 2023. "Are mathematicians, physicists and biologists irrational? Mathematical and natural science studies vs. the transitivity axiom." *Mathematical and natural science studies vs. the transitivity axiom (February 16, 2023)*.

Ramzan, M, Salman Khan, and M Khalid Khan. 2010. "Noisy non-transitive quantum games." *Journal of Physics A: Mathematical and Theoretical* no. 43 (26):265304.

Rauch, Geraldine, Kevin Kunzmann, Meinhard Kieser, Karl Wegscheider, Jochem König, and Christine Eulenburg. 2018. "A weighted combined effect measure for the analysis of a composite time-to-first-event endpoint with components of different clinical relevance." *Statistics in Medicine* no. 37 (5):749-767.

Sanjaya, Ricky, Jun Wang, and Yaodong Yang. 2022. "Measuring the non-transitivity in chess." *Algorithms* no. 15 (5):152.

Savage Jr, Richard P. 1994. "The paradox of nontransitive dice." *The American Mathematical Monthly* no. 101 (5):429-436.

Solomon, Scott D, John JV McMurray, Brian Claggett, Rudolf A de Boer, David DeMets, Adrian F Hernandez, Silvio E Inzucchi, Mikhail N Kosiborod, Carolyn SP Lam, and Felipe Martinez. 2022. "Dapagliflozin in heart failure with mildly reduced or preserved ejection fraction." *New England Journal of Medicine* no. 387 (12):1089-1098.

Steinhaus, Hugo, and S Trybula. 1959. "On a paradox in applied probabilities." *Bull. Acad. Polon. Sci* no. 7 (67-69):108.

Stohler, Michael L, and Ephraim Fischbach. 2005. "Non-transitive quantum games." *Fizika B: a journal of experimental and theoretical physics* no. 14 (2):235-244.

Temesi, József, Zsombor Szádoczki, and Sándor Bozóki. 2023. "Incomplete pairwise comparison matrices: Ranking top women tennis players." *Journal of the Operational Research Society*:1-13.

Thangavelu, Karthinathan, and Edgar Brunner. 2007. "Wilcoxon–Mann–Whitney test for stratified samples and Efron's paradox dice." *Journal of Statistical Planning and Inference* no. 137 (3):720-737.
Trybuła, Stanisław. 1961. "On the paradox of three random variables." *Applicationes Mathematicae* no. 4 (5):321-332.
Trybuła, Stanisław. 1965. "On the paradox of n random variables." *Applicationes Mathematicae* no. 8:143-156.
Богданов, Илья Игоревич. 2010. "Нетранзитивные рулетки." *Математическое просвещение* no. 14 (0):240-255.